# Ultrafast Terahertz Photoconductivity and Near-Field Imaging of Nanoscale Inhomogeneities in Multilayer Epitaxial Graphene Nanoribbons

*Arvind Singh*[1], *Jan Kunc*[2], *Tinkara Troha*[1], *Hynek Němec*[1], *Petr Kužel*[1*]

[1] *FZU - Institute of Physics of the Czech Academy of Sciences, Na Slovance 2*
*18200 Prague 8, Czech Republic*

[2] *Faculty of Mathematics and Physics, Charles University, Ke Karlovu 3,*
*12116 Prague 2, Czech Republic*

[*] kuzelp@fzu.cz

**Abstract:** We study broadband terahertz (THz) conductivity and ultrafast photoconductivity spectra in lithographically fabricated multilayer epitaxial graphene nanoribbons grown on C-face of 6H-SiC substrate. THz near-field spectroscopy reveals local conductivity variations across nanoscale structural inhomogeneities such as wrinkles and grain boundaries within the multilayer graphene. Ultrabroadband THz far-field spectroscopy (0.15–16 THz) distinguishes doped graphene layers near the substrate from quasi-neutral layers (QNLs) further from the substrate. Temperature-dependent THz conductivity spectra are dominated by intra-band transitions both in the doped and QNLs. Photoexcitation then alters mainly the response of the QNLs: these exhibit a very high carrier mobility and a large positive THz photoconductivity with picosecond lifetime. The response of QNLs strongly depends on the carrier temperature $T_c$: the scattering time drops by an order of magnitude down to $\sim 10$ fs upon an increase of $T_c$ from 50 K to $T_c > 1000$ K, which is attributed to an enhanced electron-electron and electron-phonon scattering and to an interaction of electrons with mid-gap states.

## 1. Introduction

Epitaxial graphene produced through thermal decomposition on silicon carbide (SiC) substrate is very promising for applications as it allows for the direct growth of a large-area, high-quality electronic-grade material on a semi-insulating substrate, eliminating the need for a transfer to a different substrate. [1–4] The growth on the carbon face of SiC gives rise to a van der Waals bonded multilayer epitaxial graphene (MEG) where each of the layers exhibits electronic properties similar to isolated monolayer graphene as a result of twisted, non-Bernal stacking layer arrangement. [5–9] The graphene layers closest to the substrate ("inner layers") are usually significantly doped while the rest of layers ("outer layers") is usually undoped (quasi-neutral). [10] While the supported graphene or the epitaxial graphene grown on Si-face of SiC (even after hydrogen intercalation) suffers the mobility deterioration primarily due to substrate phonon scattering, [11–15] outer layers of the graphene grown on the C-face typically display a much higher mobility due to their weak interaction with the substrate. [16–19] In turn, growing more graphene layers (which are thus quasi-neutral) scales the interaction of such layered structure with THz radiation upon optical excitation. Properties of in-plane patterned multilayer stacks on the C-face of SiC may take advantage of the high carrier mobility and of the variable Fermi energy across the layers. Indeed, this configuration leads to a high spatial field confinement, enhanced frequency tunability of the plasmonic resonance [20,21] and an increased broadband optical absorption of the stack, offering additional benefits in applications involving nonlinear graphene plasmonic properties. [22]

Apart from thorough investigations of hot carrier transport in monolayer graphene, [23–32] multi-layer graphene structures have been investigated using various optical and electronic techniques. [7,8,10,16,33–40] The inner doped layers (DLs) have been probed mostly by electronic transport measurements, [39] whereas conventional magneto-spectroscopy, far-infrared and steady-state THz spectroscopy techniques, have been employed to investigate the outer quasi-neutral layers (QNLs) of MEG. [8,38,40]

Here, we study broadband THz conductivity and ultrafast photoconductivity spectra of multilayer epitaxial graphene nanoribbons in a broad temperature range (6 – 300 K) for both polarizations of the probing THz field with respect to the ribbon direction. Modelling of these spectra reveals the relative importance of the two distinct carrier subsystems (i.e., carriers in doped and quasi-neutral layers) and enables an extraction of their key parameters such as Fermi energy and carrier mobility. Notably, the intra-band conductivity of the QNLs increases with increasing temperature on top of a nearly temperature independent contribution from the DLs and the carrier scattering time in QNLs exhibits an order of magnitude decrease as a function of carrier temperature between 50 and 1300 K. Furthermore, we employed scattering-type near-field THz microscopy (THz-SNOM) to gain insights into the local THz response and to detect variations in the conductivity at nanoscale structural features like wrinkles and grain boundaries in MEG.

## 2. Experimental details

### A. Sample preparation

We prepare the graphene films via the thermal decomposition method. Commercial Silicon Carbide (SiC, 6H polytype) wafers procured by Wolfspeed Inc were diced into 5x5 mm squares. Prior to the growth, the samples underwent a sequential cleaning process involving a sonication in acetone, followed by isopropanol, and a final rinse in deionized water to remove surface contaminants. The prepared SiC samples were then placed into a graphite crucible within a reaction chamber. The graphene growth was achieved by inductively heating the crucible to a temperature of 1650 °C for 5 minutes at 1 atm argon pressure, which helps to moderate the sublimation rate of silicon from the SiC surface, thereby facilitating the two-dimensional growth of the carbon atoms into graphene. The MEG ribbons on C-face of SiC were defined by a PMMA resist: AR-P 679.04 (Allresist GmbH) coated twice at 4000 rpm and baked at 150C/10min after each coating. The developed sample was etched in oxygen plasma (Oxford Instruments, PlasmaPro 80, 2 min, 40 mTorr, 20 sccm, 50W) from the C-face. The oxygen plasma was also used to remove graphene formed on the Si-face.

**B. Terahertz far-field spectroscopy**

Temperature-dependent (6 – 300 K) steady-state complex-valued conductivity spectra were measured using time-domain THz spectroscopy in a transmission geometry in a classical setup involving a femtosecond laser oscillator (Coherent Mira, 76 MHz repetition rate, 35 fs pulse length) [41] and reaching a frequency range of 0.15–3 THz, and in a setup employing a two-color plasma interaction in the air to generate THz pulses in a range 0.8–16 THz with the help of femtosecond regenerative amplifier (Spectra Physics Spitfire ACE, 5 kHz repetition rate, 40 fs pulse length). [42] In the following we refer to these two methods/spectral ranges as THz and multi-THz approach/spectral range, respectively. The sheet conductivity of multilayer graphene nanoribbons on SiC substrate was determined using the Tinkham formula adapted for phase-sensitive THz measurements. [43] A special care has been taken to minimize the phase error by an optimization of the substrate thickness in the conductivity retrieval procedure. [43,44]

THz photoconductivity spectra were measured at room temperature by means of a conventional optical pump–THz probe setup [45] driven by the above mentioned Ti:sapphire ultrafast amplified laser system. The sample was photoexcited at the laser fundamental wavelength of 800 nm and probed either by standard THz pulses (bandwidth of up to ~2.5 THz), or by multi-THz pulses (bandwidth of up to 16 THz). [42] We denote the measured transient spectra as $\Delta E$ (change of the transmitted THz electric field upon photoexcitation) and the reference spectra as $E^0$ (field transmitted through the unexcited sample); the sheet photoconductivity $\Delta\sigma$ is then expressed in the thin film approximation and in the small-signal limit ($\Delta E \ll E^0$) as follows [46]:

$$\Delta\sigma(\omega) = -\frac{1+N_s}{z_0}\frac{\Delta E(\omega)}{E^0(\omega)}. \quad (1)$$

Here $N_s$ ($\approx 3.14$) is the THz refractive index of SiC and $z_0$ is the vacuum wave impedance.

**C. Terahertz spectroscopy in the near field**

A THz scanning near-field optical microscope, THz-SNOM (Neaspec), was employed to investigate the local THz sheet conductivity behavior in the sample. The experimental scheme consists of an atomic force microscope, where a metallic tip (diameter ~ 50 nm) is illuminated by focused picosecond THz pulses covering a spectral range 0.5 – 2.0 THz. The common setup for steady-state THz imaging involves the THz generation and detection in InGaAs photoconductive antennas driven by femtosecond Er-doped fiber laser (Menlo systems). The nanoscale response presented in this paper is the scattered THz field demodulated at the second harmonic of the tip oscillation frequency ($\sim 50 - 100$ kHz).

## 3. Theoretical description

The total THz sheet conductivity in unpatterned MEG is given by the sum of sheet conductivities of individual graphene layers:

$$\sigma(\omega) = \sum_{j=1}^{N}\left(\sigma_j^{\mathrm{inter}}(\omega) + \sigma_j^{\mathrm{intra}}(\omega)\right) \quad (2)$$

where $N$ is the number of layers and $j$ indicates the index of the layer. The inter-band sheet conductivity of the $j$-th layer, $\sigma_j^{\mathrm{inter}}$, is connected with vertical inter-band transitions near $K$ and $K'$ points and it is given by: [28]

$$\sigma_j^{\mathrm{inter}}(\omega) = \frac{\pi e_0^2}{4h}\left[\tanh\left(\frac{\hbar\omega + 2\mu_j(T_c)}{4k_B T_c}\right) + \tanh\left(\frac{\hbar\omega - 2\mu_j(T_c)}{4k_B T_c}\right)\right] \quad (3)$$

where $\mu_j$ is the chemical potential of the $j$-th layer and $T_c$ is the temperature of charge carrier subsystem. The intra-band sheet conductivity $\sigma_j^{\mathrm{intra}}$ of the thermalized electron gas in the $j$-th layer stems from the Boltzmann transport theory:

$$\sigma_j^{\mathrm{intra}}(\omega) = \frac{1}{\pi}\frac{\tau_j}{1 - i\omega\tau_j} D_j^{\mathrm{w}}(T_c), \quad (4)$$

where the Drude weight $D_j^w$, as derived from the Boltzmann theory, reads: [47]

$$D_j^{\mathrm{w}}(T_c) = \frac{\pi e_0^2 v_F^2}{2}\int \mathcal{D}(E)\left[-\frac{\partial f_{\mathrm{FD}}(\mu_j, T_c, E)}{\partial E}\right] dE. \quad (5)$$

Here $v_{\mathrm{F}} = 1 \times 10^6$ m/s is the Fermi velocity of charge carriers, $\mathcal{D}(E)$ is the density of states of carriers on the Dirac cones, $f_{\mathrm{FD}}$ is the Fermi-Dirac distribution function and $\tau_j$ is the carrier scattering time in the $j$-th layer. We need to use this rather general integral expression since the thermal broadening of the electron distribution is comparable to the Fermi energy, particularly in the undoped layers and for elevated carrier temperatures.

The conductivity in graphene layers at zero temperature is controlled by the density of states at the Fermi energy $E_{\mathrm{F},j}$. At finite carrier temperatures $T_{\mathrm{c}}$, the broadening in the Fermi-Dirac distribution upon photoexcitation is accompanied by a decrease in the chemical potential $\mu_j$ following the carrier density conservation law [48]:

$$\frac{1}{2}\left(\frac{E_{\mathrm{F},j}}{k_{\mathrm{B}}T_{\mathrm{c}}}\right)^2 = F_1\left(\frac{\mu_j}{k_{\mathrm{B}}T_{\mathrm{c}}}\right) - F_1\left(-\frac{\mu_j}{k_{\mathrm{B}}T_{\mathrm{c}}}\right), \tag{6}$$

where $F_1$ is the first order Fermi-Dirac integral. This decrease has two important asymptotic behaviors [28] depending on the carrier temperature and the doping level:

$$\mu_j \approx E_{\mathrm{F},j}\left[1 - \left(\frac{\pi^2 k_{\mathrm{B}}^2 T_{\mathrm{c}}^2}{6E_{\mathrm{F},j}^2}\right)\right], \qquad \text{for} \quad k_{\mathrm{B}}T_{\mathrm{c}} \ll |E_{\mathrm{F},j}| \text{ (doped layers)}, \tag{7}$$

$$\mu_j \approx \frac{E_{\mathrm{F},j}^2}{4\ln(2)\, k_{\mathrm{B}}T_{\mathrm{c}}}, \qquad \text{for} \quad k_{\mathrm{B}}T_{\mathrm{c}} \gg |E_{\mathrm{F},j}| \text{ (quasineutral layers)}. \tag{8}$$

We emphasize these asymptotic relations since in our multilayer graphene sample the behaviors of the DLs and QNLs are quite different and approach either the former or the latter formula, respectively. Indeed, the low-temperature limit given by Eq. (7) applies to the DLs, where the Fermi energy significantly exceeds the thermal energy. In contrast, the high-temperature limit expressed by Eq. (8) is relevant for QNLs, where the Fermi energy is small compared to the thermal energy scale.

For steady-state measurements, the carrier temperature $T_{\mathrm{c}}$ is equal to the lattice temperature $T$; for the pump-probe measurements, $T_{\mathrm{c}}$ is elevated due to the heating of carriers by the excitation optical pulse.

It has been commonly assumed that the carrier concentration (and thus the Fermi energy) decreases approximately exponentially from the highly doped inner-most layer to a quasi-neutral outer-most layer. [10] However, as shown in the Appendix (paragraph A1), this exponential profile does not satisfactorily reproduce the temperature-dependent steady-state conductivity and transient photoconductivity spectra obtained in our measurements. Instead, we find that a model assuming two distinct types of graphene layers provides a significantly better description of our data: we thus assume in our modelling that the MEG sample is composed of $N_{\mathrm{D}}$ DLs with the Fermi energy $E_{\mathrm{F,D}}$ and the Drude weight $D_{\mathrm{D}}^{\mathrm{w}}$, and of $N_{\mathrm{QN}} = N - N_{\mathrm{D}}$ QNLs with the Fermi energy $E_{\mathrm{F,QN}}$ and the Drude weight $D_{\mathrm{QN}}^{\mathrm{w}}$. In reality, we expect that the doping profile changes more gradually, nevertheless our data strongly indicate, that the real carrier density profile is closer to a step-like transition than to an exponential one.

The total sheet conductivity of the entire MEG structure then reads

$$\sigma(\omega) = N_{\mathrm{D}}\left(\sigma_{\mathrm{D}}^{\mathrm{inter}}(\omega) + \sigma_{\mathrm{D}}^{\mathrm{intra}}(\omega)\right) + N_{\mathrm{QN}}\left(\sigma_{\mathrm{QN}}^{\mathrm{inter}}(\omega) + \sigma_{\mathrm{QN}}^{\mathrm{intra}}(\omega)\right) \tag{9}$$

(we recall that $\sigma_{\mathrm{D}}$ is the sheet conductivity of a single layer whereas we denote by $\Sigma_D \equiv N_{\mathrm{D}}\sigma_{\mathrm{D}}$ the sheet conductivity of the subsystem of all DLs; the same applies for the QNLs and we define $\Sigma_{\mathrm{QN}} \equiv N_{\mathrm{QN}}\sigma_{\mathrm{QN}}$).

Finally, we have to account for the patterning of MEG. The experimental conductivity for the probing polarization parallel to the ribbons with the period $\Lambda$ and the width $w$ then simply reflects the surface coverage $w/\Lambda$ by the ribbons,

$$\sigma_{\parallel}(\omega) = \frac{w}{\Lambda}\sigma(\omega). \tag{10}$$

For the probing polarization perpendicular to the ribbons, a localized plasmon resonance appears and the response function [43] can be derived using an equivalent circuit model consisting of a complex conductance (graphene ribbons) and capacitance (gap between ribbons) in series: [49]

$$\sigma_{\perp}(\omega) = \frac{w}{\Lambda}\frac{\sigma(\omega)}{1 + i\dfrac{w}{\Lambda}\dfrac{\sigma(\omega)}{\omega C}}; \tag{11}$$

where $C = 2\epsilon_0(1 + \epsilon_{\mathrm{SiC}})\Lambda \ln[\sec(\pi w/2\Lambda)]/\pi$ is the capacitance per unit length of the patterned graphene array. [43,50]

The formulae (10) and (11) are used for modelling temperature dependent steady-state THz spectra. We assume that the doping does not change with temperature, which means that Fermi energies $E_{\mathrm{F,D}}$ and $E_{\mathrm{F,QN}}$ as well as the numbers of doped and quasi-neutral layers $N_{\mathrm{D}}$ and $N_{\mathrm{QN}}$ are temperature independent. In the DLs, the Fermi energy is high, and the electron distribution is narrow compared to $E_{\mathrm{F,D}}$. The scattering time, $\tau_{\mathrm{D}}$, thus should not vary significantly upon cooling the sample and, in our implementation of the global fit, we assumed a single temperature independent fitted value of $\tau_{\mathrm{D}}$. This is in contrast with the carrier scattering in the QNLs: the corresponding scattering time $\tau_{\mathrm{QN}}$ is expected to be temperature dependent due to a large variation of the width of electron energy distribution compared to the Fermi energy. The fitting of temperature-dependent steady-state sheet conductivity spectra (two complex-valued spectra per temperature since we measure two polarizations) thus features 5 free global parameters, $E_{\mathrm{F,D}}$, $E_{\mathrm{F,QN}}$, $\tau_{\mathrm{D}}$, $N_{\mathrm{D}}$ and $N_{\mathrm{QN}}$ (common for all the temperatures and both polarizations), and 1 additional free parameter per temperature, $\tau_{\mathrm{QN}}$.

The differential photoconductivity spectra are modelled using the formula:

$$\Delta\sigma_{\perp,\parallel}(\omega,\tau_{\mathrm{p}},\phi)=\sigma_{\perp,\parallel}^{\mathrm{exc}}(\omega,\tau_{\mathrm{p}},\phi)-\sigma_{\perp,\parallel}(\omega)+\Delta\sigma_{\mathrm{SiC}}(\omega,\tau_{\mathrm{p}},\phi), \qquad (12)$$

where $\tau_p$ is the pump-probe delay, $\phi$ is the absorbed fluence of the pump pulse, $\sigma_{\perp,\parallel}^{\mathrm{exc}}(\omega,\tau_{\mathrm{p}},\phi)$ and $\sigma_{\perp,\parallel}(\omega)$ are the sheet conductivities in the excited and ground state, respectively, given by expressions (10) and (11) for the parallel and perpendicular geometries. The ground state conductivities $\sigma_{\perp,\parallel}(\omega)$ are represented by steady-state spectra measured at 300 K. It is also necessary to account for the differential photoconductivity of the bare SiC substrate, $\Delta\sigma_{\mathrm{SiC}}(\omega,\tau_{\mathrm{p}},\phi)$, which is non-negligible due to the high fraction of photons penetrating SiC. This contribution was measured in a separate time-resolved experiment with a bare SiC substrate.

A global fit was performed on all measured complex-valued transient spectra for both polarizations. The global parameters obtained from the fits of the temperature-dependent steady-state spectra ($E_{\mathrm{F,D}}$, $E_{\mathrm{F,QN}}$, $\tau_{\mathrm{D}}$, $N_{\mathrm{D}}$ and $N_{\mathrm{QN}}$) were kept fixed. The fit of transient datasets thus provides the carrier temperature $T_{\mathrm{c}}$ and scattering time $\tau_{\mathrm{QN}}$ per complex-valued spectrum.

## 4. Results and discussion

### A. Local probe measurements

Fig. 1(a) shows the AFM image along with the THz-SNOM image of the same area of the studied MEG sample. As deduced particularly from the scans along the lines *c*, *d* and *e* in Fig. 1(c–e), the thickness of the graphene multilayer (difference between the height in graphene regions and the height in the regions of naked substrate) is around ~5 nm. This corresponds to 14 graphene layers. While a typical MEG thickness variation in the graphene regions is ~2 nm, we observe even much higher contrasts in some regions of both the AFM and THz-SNOM images: they are mostly either due to the step bunching (i.e., a few nanometers thick terrace edge) on the substrate or due to the layer inhomogeneity during the growth itself. Epitaxial graphene is typically associated with atomic-scale defects and grain boundaries introduced during the growth process and these defects serve as scatterers that can significantly degrade the charge transport properties. [51] Indeed, such structural features like wrinkles with the size of the order of hundreds of nm and height of the order of tens of nm are clearly visible in the AFM image as demonstrated in Fig. 1(c,d,e). The wrinkles are accompanied also by contrast changes in the THz image: the wrinkles exhibit a lower conductivity [mostly thin white lines in Fig. 1(b)], which is likely due to the partly inhibited charge transport across the wrinkles.

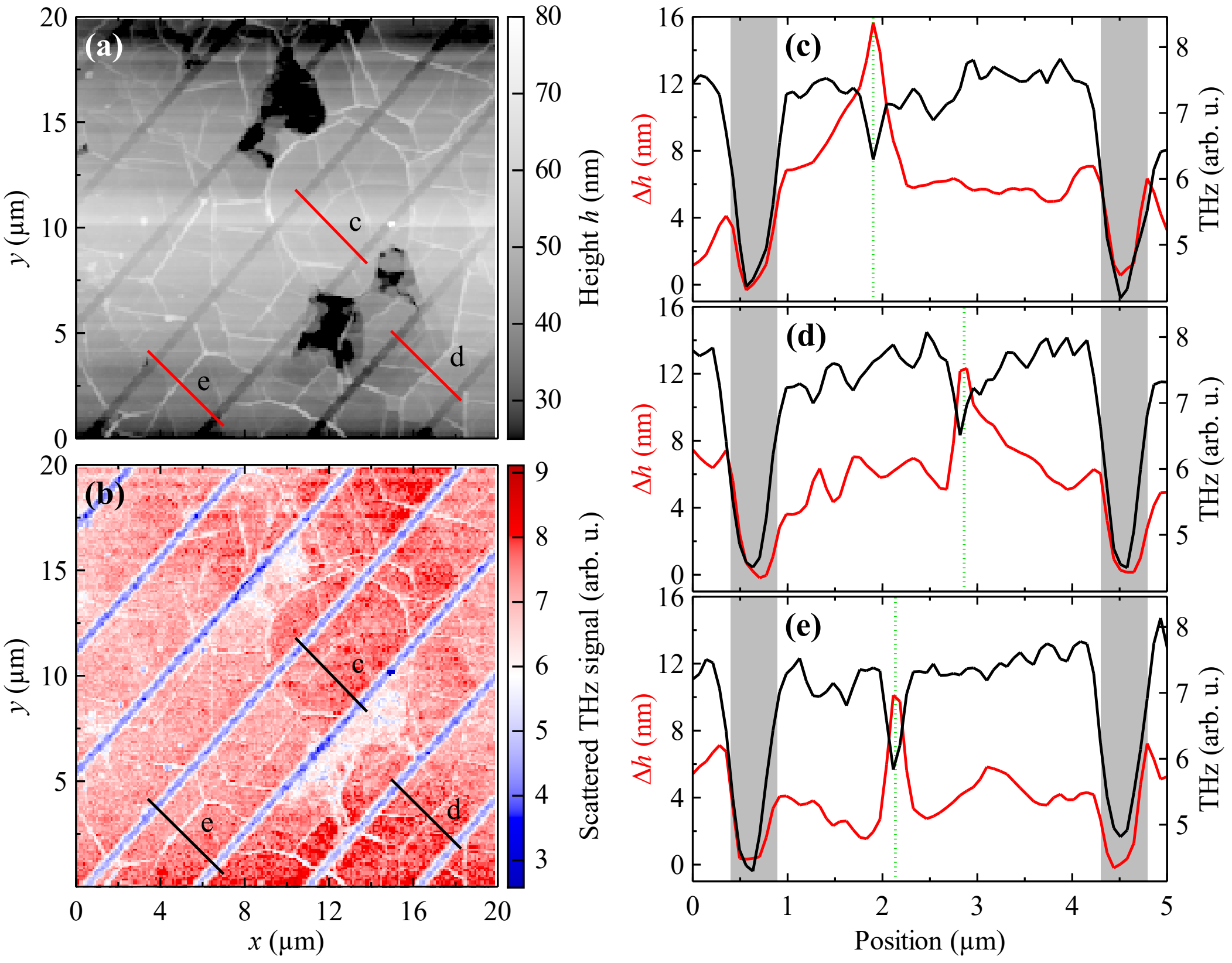

**Figure 1.** (a) AFM height profile of the studied MEG ribbons; the height changes at wrinkles are clearly observed. (b) Corresponding scattered signal from THz-SNOM, also here nanometric graphene wrinkles and grain boundaries within the MEG ribbons are resolved. (c,d,e) Scans along the lines indicated in (a) and (b) showing the thickness profile of the multilayer graphene (red curve), and the scattered THz signal at the maximum of the wave form demodulated at 2nd harmonic of the tip oscillating frequency. The grey areas indicate gaps between the graphene ribbons (note that the thickness value in the gap region is set to zero); the green dashed lines indicate the positions of graphene wrinkles or grain boundaries.

To explore the ultrafast dynamics of MEG at the nanoscale, we measured the photoinduced change in the near-field scattered THz signal in the middle of a graphene ribbon, see Fig. 2. The temporal evolution can be very well fitted by a mono-exponential decay convoluted with a Gaussian profile $G(t)$:

$$y(t) = \left[H(t - t_0) \times A_0 \exp\left(-\frac{t - t_0}{\tau_c}\right)\right] \otimes G(t) \tag{13}$$

where $H(t)$ is the Heaviside step function and the symbol $\otimes$ represents the convolution operation. The decay obtained for the lowest applied fluences agrees with the decay of the kinetics obtained in the far-field measurements (cf. red line and symbols in Fig. 2) with $\tau_D \approx$ 3.5 ps. Upon increasing the pump fluence the decay becomes somewhat slower and the amplitude $A_0$ increases and progressively saturates. This is due to a combination of two nonlinear effects: nonlinear dependence of the photoconductivity on the carrier temperature [15] and nonlinear sensitivity of the tip-enhanced THz scattering on the sample conductivity [52].

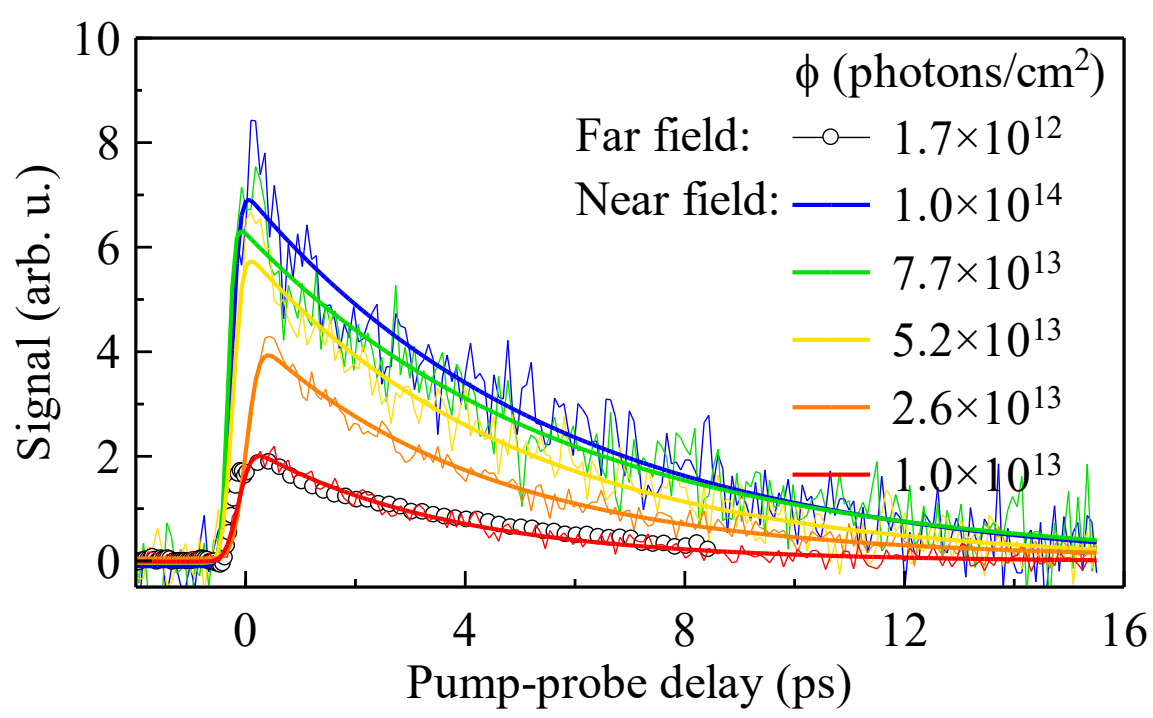

**Figure 2.** (a) Evolution of the local-probe THz SNOM signal after photoexcitation at a fixed point of the time-domain THz waveform as a function of the pump photon fluence (the absorbed fluence in graphene, $\phi$, is indicated in the legend). Thin lines represent measured data, thick lines are fits using a mono-exponential decay function convoluted with a Gaussian profile (13). The scaling of the far-field curve is arbitrary with respect to the near-field signals.

### B. Temperature-dependent steady-state conductivity

Fig. 3(a,b) shows the temperature dependence of steady-state complex sheet conductivity spectra for THz electric field parallel to the ribbons. We clearly observe that the overall conductivity increases with increasing temperature in the entire frequency range. As pointed out in Eq. (2) the total conductivity in MEG comprises intra-band and inter-band contributions in both the doped and quasi-neutral subsystems. For the DLs, $\hbar\omega \ll 2E_\mathrm{F}$ at THz frequencies and, therefore, the Pauli exclusion principle reduces the response of DLs to an intra-band Drude-like term. For the QNLs, the THz photon energies are comparable to $2E_\mathrm{F}$ and therefore both inter-band and intra-band processes may significantly contribute to the conductivity.

The temperature dependence measured in the perpendicular geometry was also involved in the global fitting procedure; however, since it does not bring any qualitatively new insight in our discussion, the data are only shown in Appendix (Fig. A2).

From the global fitting of the temperature-dependent data both in parallel (10) and perpendicular (11) geometry, we have determined that our MEG sample contains in average twelve QNLs ($N_\mathrm{QN} = 12$), with a Fermi energy $E_\mathrm{F,QN} \approx 8$ meV , and three DLs ($N_\mathrm{D} = 3$) with an average Fermi energy $E_\mathrm{F,D}$ of 200 meV.

The total number of 15 layers is entirely consistent both with the measured optical absorption of the sample reaching 22% at 800 nm and with the AFM profile of the ribbon-patterned sample (see Fig. 1). We used the transfer matrix formalism to verify that 22% of the incident pump power is indeed absorbed in the studied MEG stack composed of 15 layers and that, in this case, the interferences inside MEG ensure an almost uniform excitation depth profile.

The low number of the “doped” layers is consistent with the findings in Ref. [10]: here we are unable to resolve and disentangle the response of each individual doped layer, we sense just an average response of the doped subsystem. On the other hand, unlike in Ref. [10], THz spectral range is rather sensitive to the presence and properties of the quasi-neutral layers; in particular, we can conclude that there is an elevated number of layers with the Fermi energy significantly below the thermal energy at room temperature. The determined value of 8 meV for the Fermi energy in QNLs is in a very good agreement with the results obtained previously. [38,53]

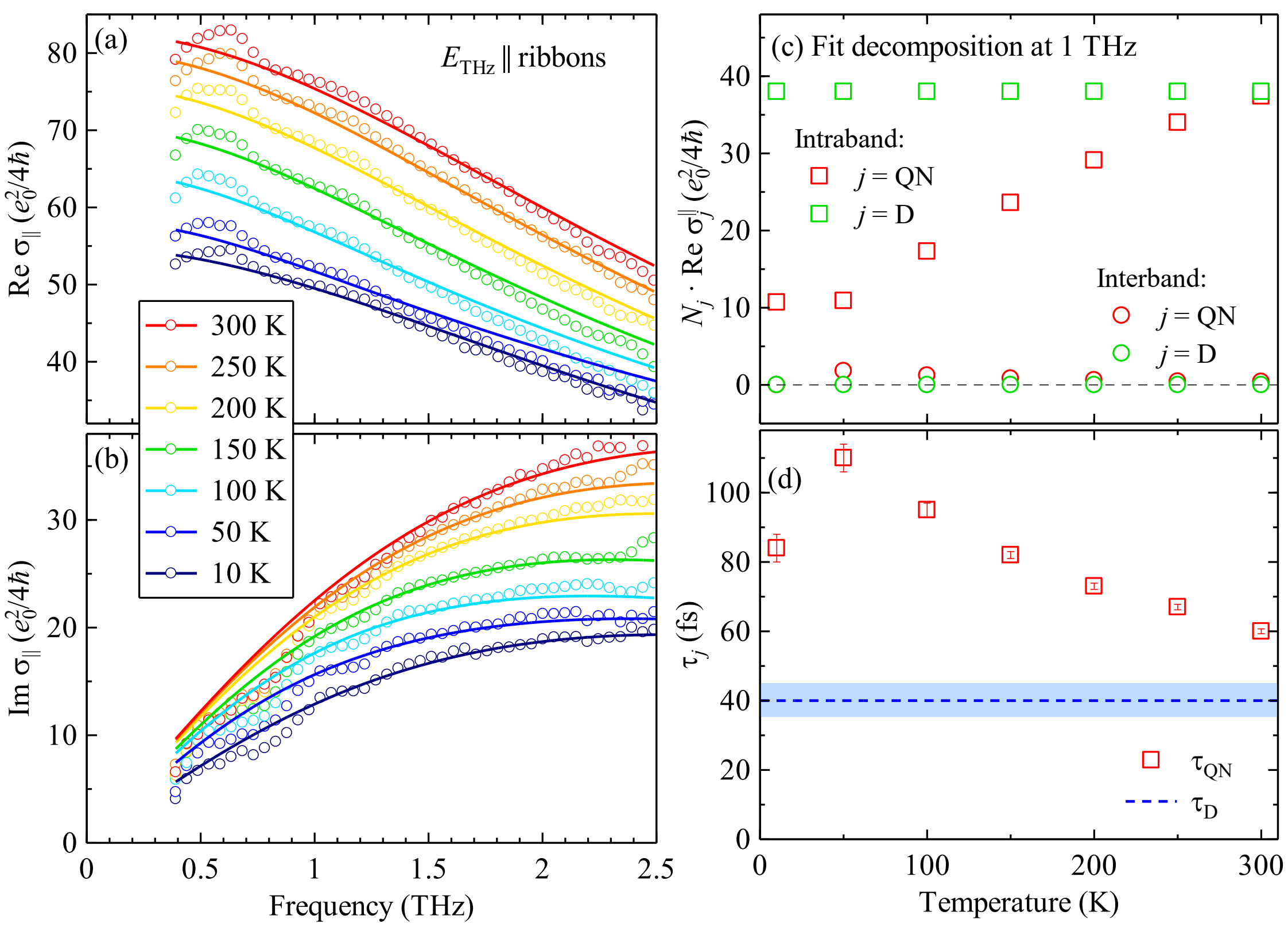

**Figure 3.** Real (a) and imaginary (b) parts of the steady-state terahertz sheet conductivity of MEG ribbons as a function of the sample temperature; $E_{\mathrm{THz}} \parallel$ ribbons. Symbols: measured data; lines: fits to the measured data using the model (10). (c) Temperature dependences of the intra-band (squares) and inter-band (circles) sheet conductivity contributions from the ensemble of the DLs (green) and QNLs (red) at 1 THz. (d) Temperature variation of the scattering time $\tau_{\mathrm{QN}}$; the scattering time $\tau_{\mathrm{D}}$ was a global temperature-independent parameter in the fits and we show its value and uncertainty.

Fig. 3(c) shows a decomposition of the signal into individual contributions: inter-band and intra-band contributions from QNLs and also these contributions from DLs. The inter-band conductivity in the DLs completely vanishes and the intra-band conductivity in the DLs remains almost temperature independent, see Fig. 3(c), since the corresponding Fermi energy of $\sim 200$ meV largely exceeds the energy of thermal fluctuations. Hence, the observed temperature change of the total conductivity between 10 and 300 K is entirely due to the variation of the contributions originating in the QNLs, see Fig. 3(c). At 10 K the inter-band contribution in QNLs vanishes since the states are fully occupied up to $\sim 8$ meV ($\sim 2$ THz); at higher temperatures this contribution becomes non-zero since the thermal broadening ensures that states are available for inter-band transitions in the THz range but, anyway, it does not dominate. For the temperatures above 50 K the number of valence-band electrons near the Dirac point become reduced and, progressively more electrons become available for Drude-like transport upon the temperature increase; therefore, the intra-band conductivity dominates and strongly increases with temperature.

As evidenced by Fig. 3(d), the temperature variation is accompanied by changes in the scattering time in the QNLs. Several mechanisms can contribute to the observed significant decrease of the scattering time upon temperature increase above 50 K. ($i$) The low carrier density in QNLs weakens the electronic screening. As a result, electrons in these layers experience stronger Coulomb interactions with each other, enhancing electron-electron scattering at a higher electron thermal energy. Also, the weak screening makes the electrons more sensitive to lattice vibrations, so that electron-phonon scattering becomes increasingly efficient at higher temperatures. ($ii$) Previous theoretical studies [54,55] supported by the experimental results of S. Massabeau et al., [38] have shown the existence of a finite density of states (often referred to as mid-gap states) at the Dirac point, arising from vacancies and interface states in the QNLs. At higher temperatures, more carriers can be thermally activated into or out of these states, providing additional scattering channels (carrier trapping/releasing, etc.).

The retrieved scattering time drops also upon cooling from 50 down to 10 K. In classical semiconductors a similar behavior is usually observed, and it is attributed to an interaction of carriers with the charged impurities; we can hypothesize that a similar process may occur also in our case.

In the DLs the situation is quite different. On the one hand, the higher carrier density provides a stronger screening, which suppresses electron-electron and electron-phonon scattering. On the other hand, random potential fluctuations caused by impurities in the substrate introduce additional disorder, which strongly reduces the scattering time in the DLs. [56] As a result, the scattering time in the DLs ($40 \pm 5$ fs) is shorter compared to the QNLs.

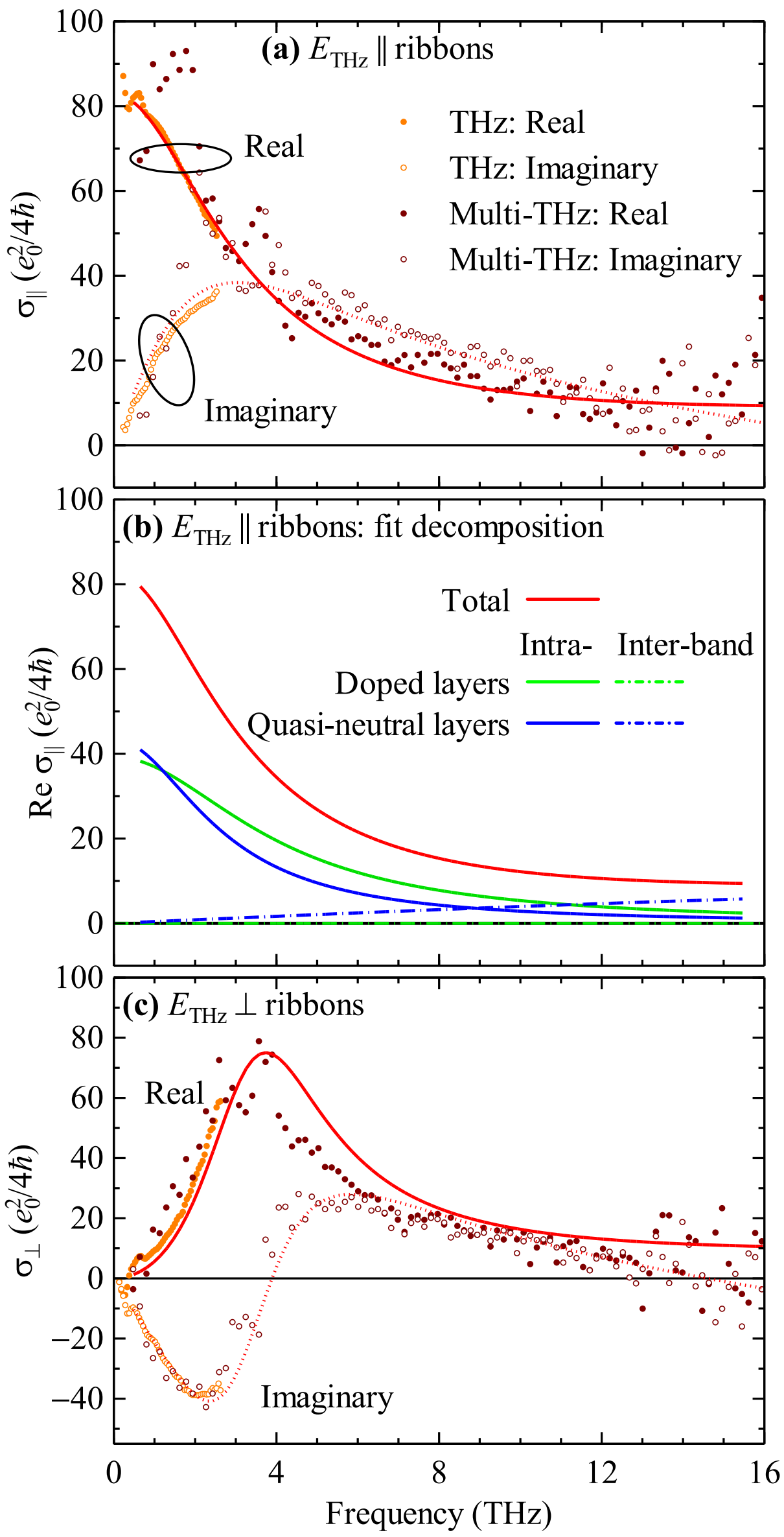

**Figure 4.** Measured THz and multi-THz complex sheet conductivity spectra in the steady state at room temperature for both polarizations of the probing field. (a,c) THz electric field parallel (a) and perpendicular (c) to the ribbons. Lines: fits with Eq. (10) or (11) for parallel or perpendicular geometry, respectively. For fitting the data from the standard THz range were only used (orange symbols), whereas the multi-THz data (dark red symbols) were not involved in the fitting procedure. (b) Decomposition of the fit into individual components for the parallel geometry.

This scattering time is also somewhat shorter than that observed in a similarly doped quasi-free-standing single layer graphene grown on Si face of SiC substrate (49 ± 1 fs for the polarization parallel to the ribbons [43]). The decrease of the scattering time correlates with the significantly worse homogeneity of MEG (Fig. 1): both wrinkles and grain boundaries are likely to act as scattering centers reducing the scattering time and the carrier mobility. It should be emphasized that these reduced values represent effective properties of graphene averaged over the probed macroscopic area in the far-field approach.

The carrier mobilities $\eta_{\mathrm{D}}$ and $\eta_{\mathrm{QN}}$ for the two electronic subsystems can be calculated using the expression: [43]

$$\eta_j = \tau_j \, e_0 v_{\mathrm{F}}^2 / \mu_j(T) \tag{14}$$

and we find $\eta_{\mathrm{D}} \approx 2300\ \mathrm{cm^2V^{-1}s^{-1}}$ and $\eta_{\mathrm{QN}} \approx 7.5 \times 10^5\ \mathrm{cm^2V^{-1}s^{-1}}$ at room temperature in the DLs and QNLs, respectively. The observed ultrahigh carrier mobility close to the Dirac point is consistent with the values predicted in the literature [57] as well as with the values obtained from the Landau level spectroscopy. [35] The small width of the ribbons may impose questions about the validity of the Drude model for high-mobility charge carriers. In general, localization effects in the conductivity spectra appear when the nanostructure size is comparable to or smaller than the diffusion length during a single period of the lowest probing radiation frequency. [25] The diffusion coefficient can be determined from the Einstein relation generalized for the linear band dispersion: [58] $D_{\mathrm{dif}} = \frac{\eta_j}{e_0} \frac{\partial n}{\partial \mu_j} \approx \frac{\tau_j v_{\mathrm{F}}^2}{2}$. The corresponding diffusion length, $\sqrt{D_{\mathrm{dif}}/f}$, is then 230 and 350 nm for $f = 0.4$ THz in DLs and QNLs, respectively. Since the nanoribbons are about 10 times wider, the localization effects can be safely neglected.

Fig. 4 shows the complex sheet conductivity spectra in a frequency range extended up to 16 THz at room temperature for both polarizations of the probing THz field. Note that the fits were performed on the set of the standard THz data only as described above (orange symbols up to 2.8 THz for the room temperature in Fig. 4a,c) and, subsequently, high-frequency extrapolations of these fits were compared to the multi-THz spectra shown here. An excellent match is observed for both polarizations between the high-frequency behavior of the fitting curve and the experimental multi-THz data confirming the validity of the fitting model in an ultrabroad frequency range. In the case of $E_{\mathrm{THz}} \perp$ ribbons a full resonance curve is observed with a significant plasmonic conductivity peak at $\sim 4$ THz. The plasmonic behavior is associated with the separation of charge carriers within the ribbons. Fig. 4(b) displays a decomposition of the fit into individual components for the parallel geometry. The intra-band transport clearly dominates and the contributions from the ensembles of DLs and QNLs are comparable. A non-vanishing inter-band contribution from QNLs is also observed in the high-frequency part of the spectrum and it even exceeds the intra-band contributions in the upper-most frequency range. Similar decomposition is not presented for the perpendicular geometry since the individual contributions are not additive owing to the resonant interaction expressed by Eq. (11).

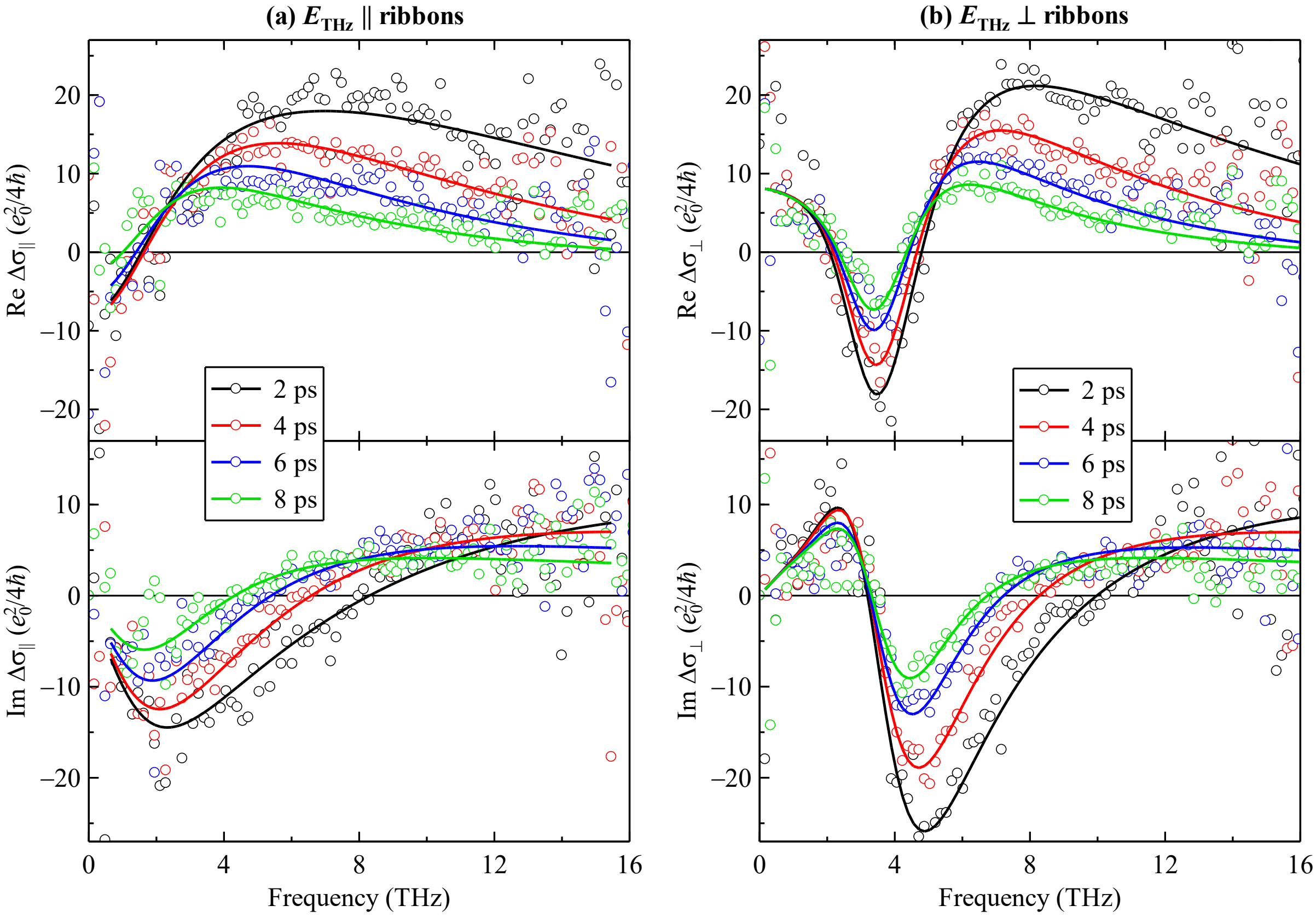

**Figure 5.** Differential sheet photoconductivity spectra in parallel (a) and perpendicular (b) configurations (absorbed pump photon fluence is $1.9 \times 10^{14}$ photons $cm^{-2}$, temperature is 300 K). Symbols: measured data, lines: results of the global fit described in the text.

Compared to our previous work on Si-face-grown quasi-free-standing single layer graphene (QFSLG) with a similar lithographically prepared ribbon structure (namely with the same spatial width and period) [43], the plasmonic resonance here is blue shifted by about a factor of two and it appears at ~ 4 THz. This is due to a nearly four times higher total conductivity of the graphene structure grown on the C face, which is caused by the higher number of graphene layers in the sample investigated here.

### C. Ultrafast photoconductivity

Fig. 5 shows differential sheet photoconductivity spectra for THz electric field parallel (a) and perpendicular (b) to the ribbons for several pump-probe delays. To discuss more in depth these results, we present in Fig. 6(a) the decomposition of the sheet photoconductivity obtained in the parallel geometry for a pump-probe delay of 2 ps into inter- and intra-band contributions in the DLs and QNLs. (Note that for the perpendicular geometry the above-mentioned individual contributions are not additive due to the resonant plasmonic behavior given by Eq. (11); therefore they are not shown here.) We remark that, particularly at low frequencies, the "background" Drude-like photoconductivity of the SiC substrate is not negligible and we have to take it into account in our analysis.

According to Fig. 6(a), the sheet photoconductivity of MEG in the parallel geometry is clearly dominated by intra-band transitions in the QNLs across the entire frequency range. Although its spectral shape resembles a broad resonance at non-zero frequency, it is in fact given by a difference between the respective Drude contributions in the photoexcited and ground state: the photoexcited state is characterized by a higher Drude weight and shorter scattering time compared to the ground state. The increase of the Drude weight upon photoexcitation is due to the excess population of photocarriers in the conduction band in the QNLs. This behavior is indeed captured in our model: upon photoexcitation the chemical potential in QNLs shifts further towards the neutrality point [$\mu_{\mathrm{QN}} \to 0$ following Eq. (8)], and Eq. (5) then yields an approximately linear increase of the Drude weight (i.e., of the photoconductivity) with the carrier temperature. This is accompanied by a decrease of the scattering time resulting from carrier heating [Fig. 7(b)].

Further contributions to the sheet photoconductivity of MEG in the parallel geometry are marginal. Upon photoexcitation the Drude weight in DLs slightly decreases, since the excess photocarriers merely heat up the existing subsystem of equilibrium carriers and the chemical potential does not change much compared to its equilibrium value [Eq. (7)], which leads to a barely visible photoconductivity signal. The inter-band sheet photoconductivity of QNLs is weakly negative and its magnitude monotonically increases with increasing frequency (Fig. 6a): this is attributed to an increased Pauli blocking at higher frequencies upon photoexcitation (indeed, after photoexcitation the hot carriers partially fill the states for inter-band transitions), and it reflects the carrier energy distribution during the cooling process after photoexcitation.

The sheet photoconductivity signal for $E_{\mathrm{THz}} \perp$ ribbons has a rather complex shape as observed in Fig. 5(b). We show in Fig. 6(b) an example of a decomposition of the complex-valued transient spectrum to the excited-state conductivity, ground-state conductivity and SiC background photoconductivity: these spectra correspond to the individual terms at the right-hand-side of Eq. (12). We note in Fig. 5(b) a weak non-zero contribution to the real part even at the lowest frequencies: this feature is entirely due to the Drude photoconductivity of the SiC substrate, since—according to Eq. (11)—the effective response of the ribbons vanishes upon $\omega \to 0$. Besides that, the shape of the transient signal is determined by the difference between the response of plasmons in the excited and ground state. In the ground state the contributions of DLs and QNLs to the plasmon resonance have approximately the same magnitude. In the excited state, the DLs and QNLs influence the spectral position of the plasmon in an opposite way. Heating of carriers in DLs contributes to a plasmon red shift, however this contribution is weak. More importantly, the plasmon is strongly influenced by the large amount of excited hot carriers in QNLs, leading to a significant increase in the plasmonic frequency. The resulting plasmon is thus blue shifted and also broadened due to the opposite contribution of QNLs and DLs and due to an increased carrier temperature. This is in contrast with the behavior of the plasmon in QFSLG [43] where the single graphene layer is strongly doped and, consequently, the plasmon exhibits a weaker red shift upon photoexcitation.

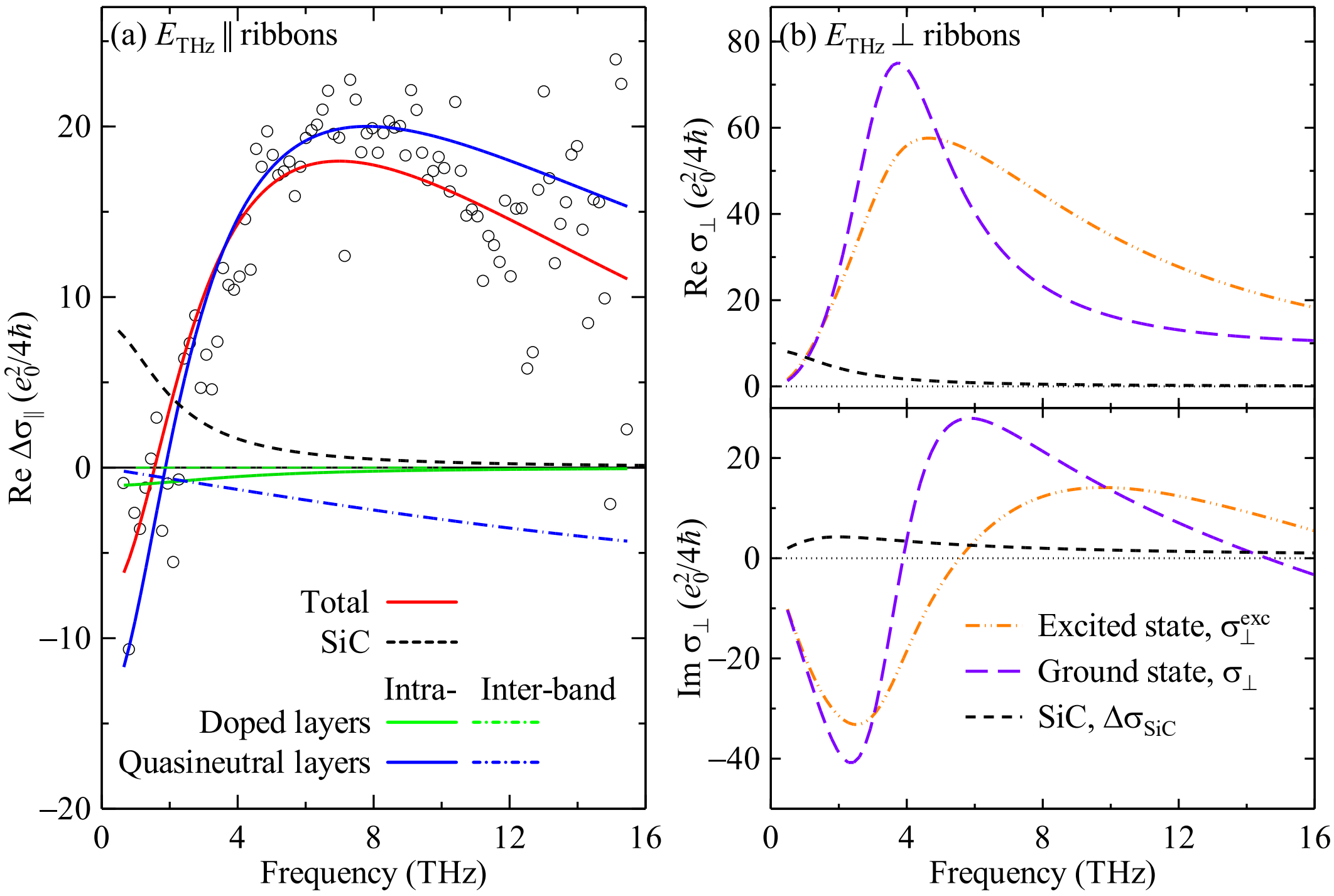

**Figure 6.** (a) Decomposition of the sheet photoconductivity spectra at 2 ps for $E_{\mathrm{THz}} \parallel$ ribbons into inter- and intra-band contributions in the DLs and QNLs following Eq. (9). (b) Decomposition of the sheet photoconductivity spectra at 2 ps for $E_{\mathrm{THz}} \perp$ ribbons into the sheet conductivities in the excited state, ground state and the contribution of SiC following Eq. (12).

In Fig. 7(a,b), we compile the evolution of the fitted parameters. At 2 ps after photoexcitation, the carrier temperature is still rather high (1300 K) and it recovers with a decay time of about 3 ps. This time scale agrees with the decay observed using the local probe spectroscopy shown in Fig. 2. The scattering of hot carriers is significantly enhanced with increasing carrier temperature: compared to the room temperature, the scattering time in the QNLs drops by a factor of 4 to 5 to $\sim 10$ fs for $T_c > 1000$ K, see Fig. 7(b). This prominent decrease is attributed to the enhanced carrier scattering by optic phonons since the excess energy of hot carriers exceeds the optical phonon energy. [28] The decrease is even stronger than that observed in QFSLG in our previous work. [43] Since QNLs contain very few carriers at equilibrium, its optical excitation represents quite a large perturbation. The combination of weaker screening, larger available phase space, and efficient electron-phonon interaction can therefore be expected to result in a more pronounced decrease in scattering time compared to a single doped layer in QFSLG.

The temperature dependence of the carrier mobility calculated using Eq. (14) is compiled in Fig. 7(c). A slow increase of the mobility with temperature due to a decreasing chemical potential is observed in DLs. The carrier mobility in QNLs first increases for the same reason; however, at higher temperatures, the decrease of the chemical potential is compensated by the decrease of the carrier scattering time and the mobility thus remains almost constant (see also Appendix A3 for the temperature dependence of the chemical potential).

Finally, we wish to justify our implicit assumption of the same carrier temperature throughout all layers of the MEG sample. A similar assumption was also used previously in all-optical pump-probe experiments with MEG [59]. However, more recently a doping-dependent cooling rate (between 2.5 ps and 4 ps) has been observed in an exfoliated graphene encapsulated between hexagonal boron nitride layers [60]. In this study the in-plane carrier cooling was found inefficient due to a long momentum scattering time ($\tau \sim 400$ fs) and the cooling process was dominated by an out-of-plane heat transfer towards encapsulating materials. In contrast, in the MEG samples the momentum scattering time is clearly sub-100 fs at (and above) the room temperature and the cooling thus occurs predominantly by a disorder-enabled mechanism of scattering on acoustic phonons [15]. Moreover, our transient signal is nearly entirely due to the electrons in QNLs (see Fig. 6) and a small difference in the cooling rate in DLs does not influence the signal. Another issue concerns interlayer energy transfer in MEG reported in Ref. [37]. This effect is associated with transient carrier temperature differences among layers which is rather small in our case and the process occurs on a tens of picoseconds timescale which is considerably slower than the picosecond dynamics in our sample. The low importance of these effects is further confirmed by our observation of an almost identical cooling rate in QFSLG [43] as in our sample.

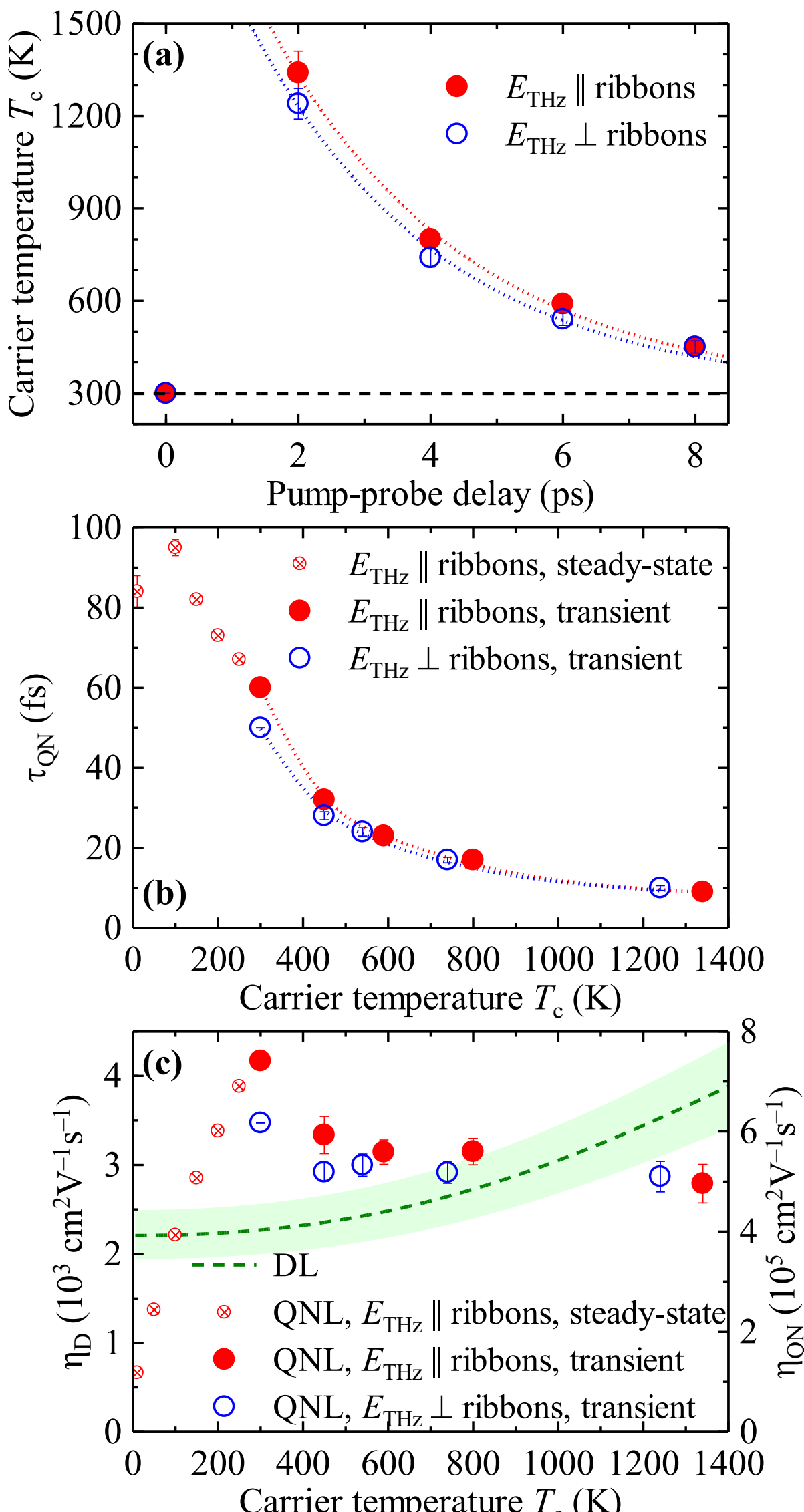

**Figure 7**. (a) Evolution of the carrier temperature in the QNLs after photoexcitation. The dashed line indicates the room temperature; the dotted lines serve only to guide the eye. (b) Compilation of the scattering times from transient and steady-state measurements versus carrier temperature. (c) Compilation of the carrier mobilities from transient and steady-state measurements versus carrier temperature calculated using Eqs. (6) and (14).

## 5. Conclusion

We study broadband THz conductivity response of multilayer epitaxial graphene nanoribbons in a large lattice temperature range (10 – 300 K) for both polarizations of the probing THz field with respect to the ribbons. We find that the multilayer can be well described as a two-component system of doped inner layers and quasi-neutral outer layers. As the sample temperature rises from 6 to 300 K, the intra-band conductivity of the quasi-neutral layers increases on top of a nearly constant contribution from the doped layers. The ultrafast photoconductivity response is mainly governed by a light-induced increase of the intra-band conductivity in the quasi-neutral layers, whereas the contribution of doped layers is negligible. The outer layers exhibit exceptionally high carrier mobility of $7.5 \times 10^5\ \mathrm{cm^2V^{-1}s^{-1}}$, and their conductivity can be controlled on picosecond timescales via optical excitation. Our ultrabroadband approach involving probing frequencies over two decades allows us to disentangle the intra- and inter-band carrier transport mechanisms and quantify their interplay, demonstrating a strong temperature dependence of the carrier scattering in the quasi-neutral layers. Near-field THz spectroscopy provides complementary insights into the multilayer structure, revealing local variations in conductivity across nanoscale inhomogeneities such as wrinkles and grain boundaries, which are otherwise unresolved in far-field measurements.

**Acknowledgment**

The authors acknowledge financial support by Czech Science Foundation (grant number 24-10331S). CzechNanoLab project LM2023051 funded by MEYS CR is acknowledged for the financial support of the sample fabrication at CEITEC Nano Research Infrastructure. This work was also supported by European Union and the Czech Ministry of Education, Youth and Sports (Project TERAFIT– CZ.02.01.01/00/22_008/0004594).

**Data availability**

The data that support the findings of this article are openly available [61].

## APPENDIX

### A1. Exponential decay model of the carrier density across layers

Epitaxially grown multilayer graphene is electron-doped through interaction with the underlying SiC substrate, resulting in a rapid decrease of n-type carrier density with the distance from the substrate. Following Ref. [10], we also tried to model the reduction of the carrier density or, equivalently, of the Fermi energy as an exponential decay

$$E_{\mathrm{F},j} = E_{F,1} \exp\left(-\frac{(j-1)}{2L_s}\right). \qquad \text{(A1)}$$

where $j$ is the layer index ($j = 1 \dots N$) and $L_s$ denotes the charge screening length. For simplicity, we further consider that the scattering time decreases linearly with the Fermi energy:

$$\tau_j = \tau_1 + \Delta\tau(T_c)\left[1 - \frac{E_{\mathrm{F},j}}{E_{F,1}}\right], \tag{A2}$$

where $\Delta\tau(T_c)$ is a fitted temperature dependent change of the scattering time upon doping. For fitting the complex THz spectra in the parallel geometry, a global fit was performed. The best agreement with experimental data was obtained for $N = 20$ (shown in Fig. A1), with the global fit parameters (common for all the spectra): $E_{F,1} = 160$ meV, $\tau_1 = 50$ fs and $L_s = 1.7$. However, we observe systematic deviations of the fit from the experimental data, which could not be removed within the exponential carrier decay model. In addition, the total number of layers obtained from the fit does not match the total film thickness observed using AFM. We thus conclude that the step-like change of the Fermi level employed in the text is a much better fitting model than the exponential Fermi-level depth-profile.

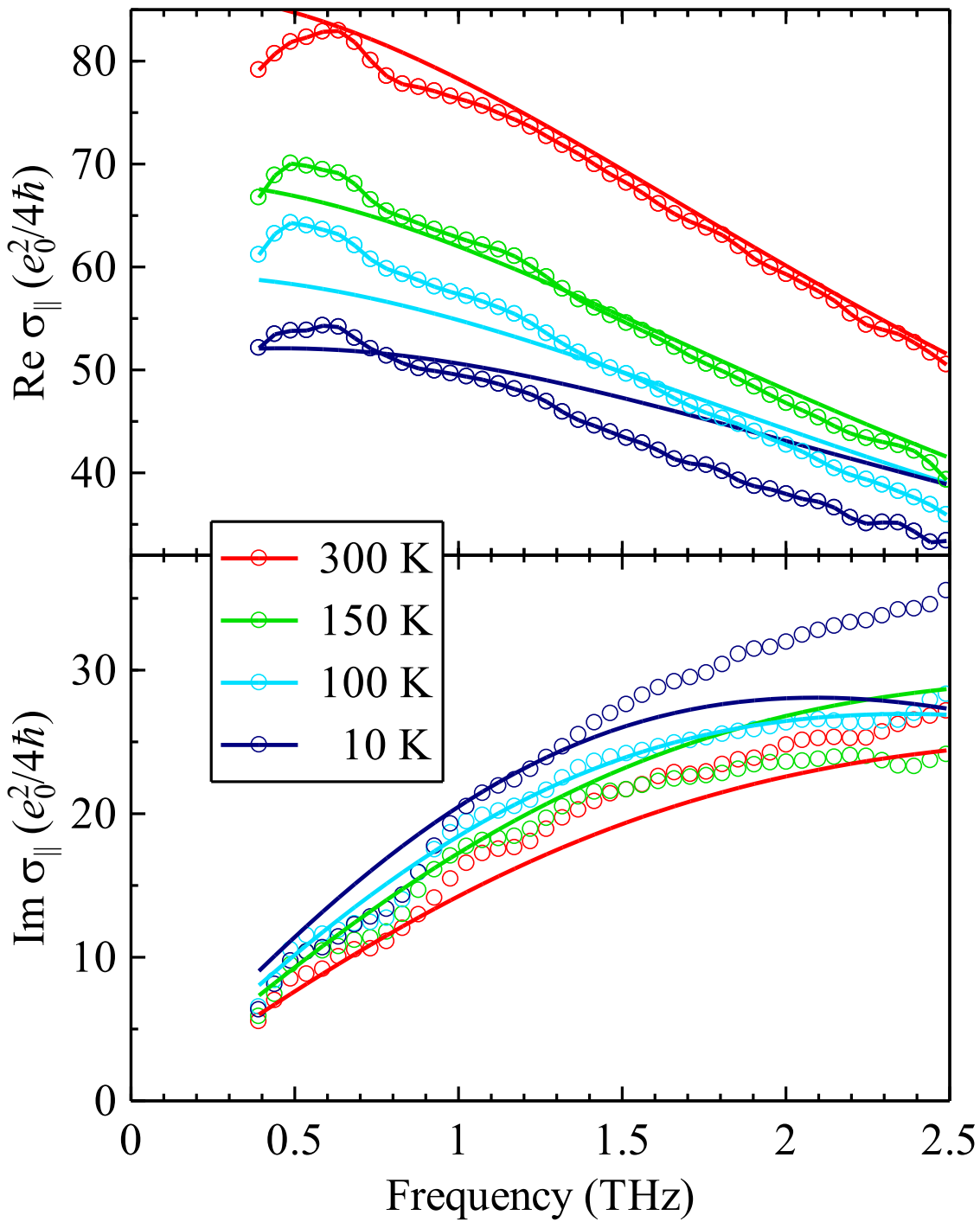

**Figure A1.** Selected temperature-dependent steady-state sheet conductivity spectra of MEG sample for $E_{\mathrm{THz}} \parallel$ ribbons. Symbols: measured data; lines: best fit to the measured data considering an exponential distribution of the carrier density ($N = 20$).

### A2. Temperature-dependent steady-state conductivity for $E_{\mathrm{THz}} \perp$ ribbons

In Fig. A2 we display the measured and fitted steady-state sheet conductivity spectra of MEG sample for the THz field perpendicular to the ribbons.

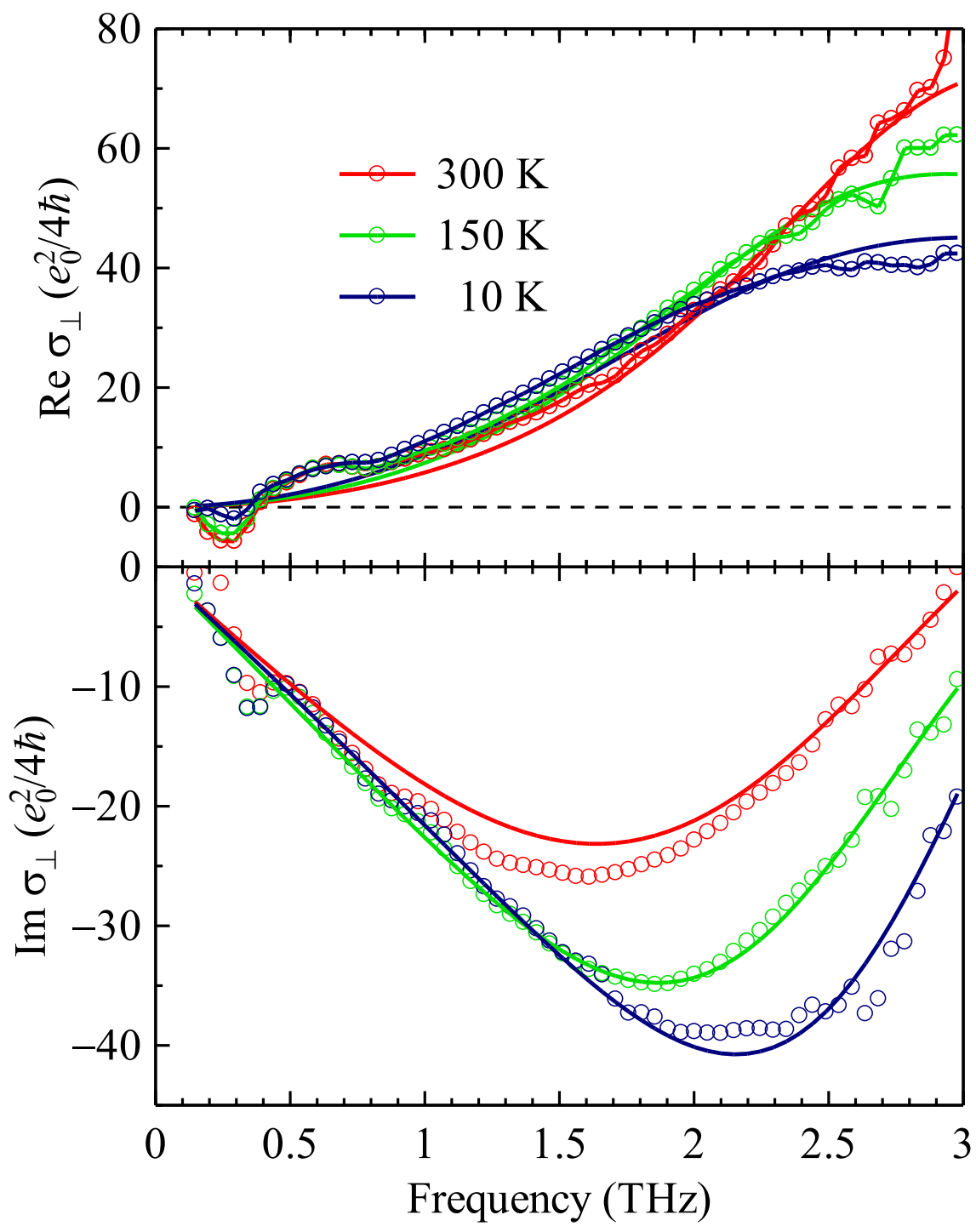

**Figure A2.** Selected temperature-dependent steady-state sheet conductivity spectra of MEG sample for $E_{\mathrm{THz}} \perp$ ribbons. Symbols: measured data; lines: fits to the measured data using the model presented in the main paper given by Eqs. (9) and (11).

### A3. Temperature dependence of the chemical potential

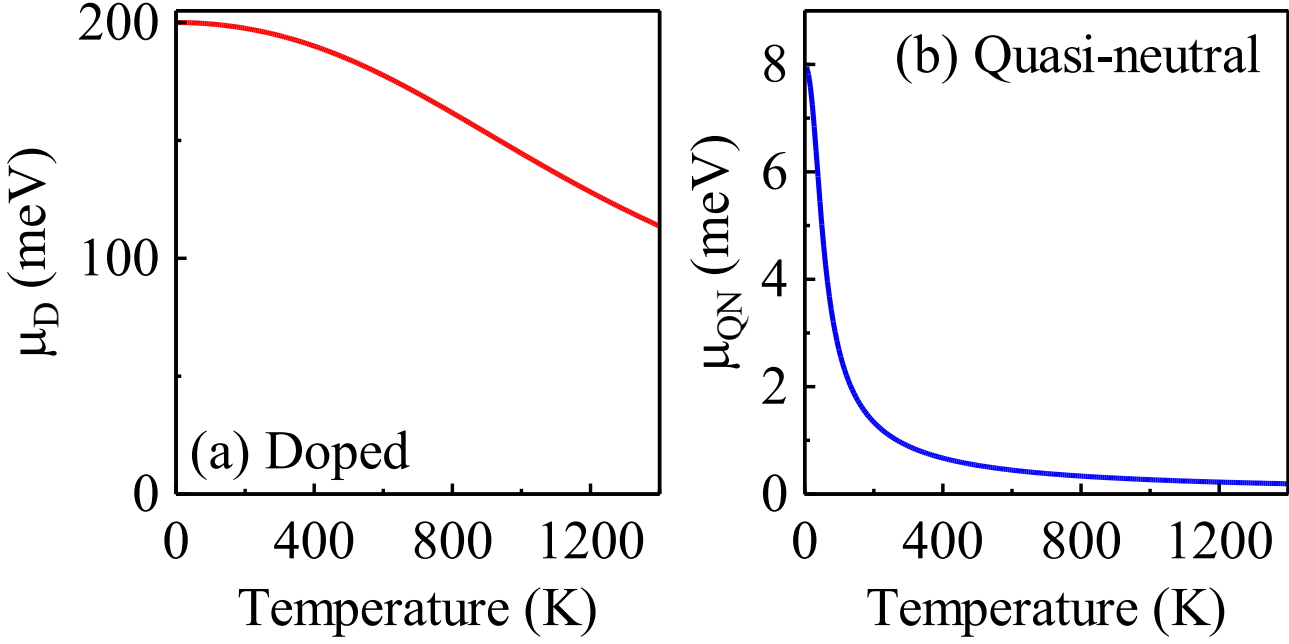

**Figure A3.** Temperature dependence of chemical potentials of (a) DLs with $E_{\mathrm{F,D}} = 200$ meV and (b) QNLs with $E_{\mathrm{F,QN}} = 8$ meV.

In Fig. A3 we display the temperature dependence of chemical potentials calculated using Eq. (6).